\documentclass{article}

 \usepackage{graphicx} \usepackage{amsmath}

\suppressfloats[t]

\begin{document} \large

\begin{center}
{\bf Formation of Weak Singularities on the Surface of a Dielectric Fluid in a Tangential Electric Field}\\[1.0ex]

\vspace{2mm}
{\it Evgeny A. Kochurin$^1$ }\\[1.0ex]
{$^1$Institute of Electrophysics, Ural Division,
Russian Academy of Sciences, 106 Amundsen Street, 620016 Ekaterinburg, Russia\\
}
\end{center}

\vspace{2mm}
\begin{abstract}

\large
The process of interaction between nonlinear waves on a free surface of a nonconducting fluid in a strong tangential electric field is simulated numerically (effects of the force of gravity and capillarity are neglected). It is shown that singular points are formed at the fluid boundary in a finite time; at these points, the boundary curvature significantly increases and undergoes a discontinuity. The amplitude and slope angles of the boundary remain small. The singular behavior of the system is demonstrated by spectral functions of the fluid surface -- they acquire a power dependence. Near the singularity, the boundary curvature demonstrates a self-similar behavior typical for weak root singularities.

\end{abstract}

\vspace{2mm}

Describing the nonlinear dynamics of fluid interfaces is an extremely difficult problem for an analytical investigation. In many nonlinear models of the motion of fluids with free or contact interfaces, there can appear the so-called weak singularities the formation of which is accompanied by generation of regions with infinite curvature at the fluid interface. For such singularities, the amplitude and steepness of the interface (slope angles) remain small. Weak singularities can appear with the development of the Kelvin-Helmholtz instability [1, 2], Tonks-Frenkel instability [3,4], and even during inertial motion of a fluid with a free boundary [5] (the force of gravity and capillary forces are not included into the consideration). It is important that the description of weak singularities does not break criteria of applicability of weakly nonlinear models allowing one to obtain exact solutions of motion equations based on the method of characteristics. Another type includes strong singularities for which the amplitude or steepness of the boundary undergo a discontinuity. Examples of such singularities are strongly nonlinear structures (cusps, cones, angles, etc.) forming, e.g., on the fluid surface in the field of gravity [6], at fluid boundaries in a strong electric field [7-9], as well as in situations in which the fluid boundary moves without acceleration [10, 11].

Boundary hydrodynamic instabilities can be suppressed by an external electric field directed tangentially to the unperturbed boundary of nonconducting fluids [12]. This property of the horizontal electric field is associated with the considerable interest in studying its influence on the nonlinear dynamics of fluid boundaries (see [13-16]). In [17-19], exact partial solutions of motion equations for a high-permeability dielectric fluid placed in a strong horizontal electric field were found (the effect of capillary and gravitational forces is negligible). These solutions describe the propagation of nonlinear surface waves without distortions in the direction of the external field or oppositely to it. Numerical simulation of strongly nonlinear dynamics of the fluid surface demonstrates that the interaction of counter-propagating waves leads  to the formation of singular points at the boundary where the field strength and fluid velocity undergo a discontinuity and the surface curvature unboundedly increases [20]. The formation of such discontinuities is accompanied by the overturn of surface waves (the slope angles tend to $\pi$/2), which impedes the analytical description of the fluid boundary.

In this work, the dynamics of the nonconducting fluid surface in a tangential electric field is considered based on a simpler model, namely, a weakly nonlinear one. Below, it is shown that the fluid boundary during the interaction of counter-propagating waves demonstrates the behavior which is very close to the formation of weak singularities observed in different physical situations [1–5]. This fact testifies to the integrability of complete electrohydrodynamics equations of fluids with a free boundary in a strong tangential electric field.

Let us consider the potential flow of a deep incompressible ideal dielectric high-permeability fluid placed in an external homogeneous horizontal electric field. Due to the presence of anisotropy related to a distinguished direction of the electric field in the problem under study, we consider only plane-symmetric surface waves propagating parallel to the external field. We suppose that the field strength is directed along the $x$ axis (the $y$ axis of the Cartesian coordinate system is directed along the normal to it) and is equal to $E$ in the absolute value. Function $\eta(x, t)$ specifies the boundary deviation from the unperturbed state $y = 0$. The dispersion relation for linear waves at the boundary has the form [17-19]
$$\omega^2=c^2k^2,\qquad c=\sqrt{\frac{\varepsilon_0 \varepsilon }{\rho}} E,$$
where $\omega$ is the frequency; $k$ is the modulus of the wave vector; $\varepsilon_0$ is the electric constant; $\varepsilon$ and $\rho$ are the permittivity and density of the fluid, respectively ($\varepsilon\gg1$); and $c$ is the speed of wave propagation. For the further analysis, it is convenient to introduce dimensionless variables:
$$y\to y/k_0,\quad x\to x/k_0, \quad t\to t/k_0 c,$$
where $k_0$ is the characteristic value of the wave number. In the dimensionless variables, the wave propagation speed is equal to unity.

The equations of the boundary motion up to quadratically nonlinear terms have the form [17]
\begin{equation}\label{eq1}\psi_t=\hat H \eta_x +\hat H(\eta \hat H \eta_x)_x+\partial_x(\eta \eta_x) +\frac{1}{2}(\hat H \eta_x^2-\eta_x^2) +\frac{1}{2}(\hat H \psi_x^2-\psi_x^2),\end{equation}
\begin{equation}\label{eq2}\eta_t=-\hat H \psi_x- \hat H(\eta \hat H \psi)_x-\partial_x(\eta \psi_x),\end{equation}
where $\psi$ is a function determining the fluid velocity potential at the boundary and $\hat H $ is the Hilbert transform; its action in the Fourier space is defined as $\hat H f_k=i \mbox{sign}(k) f_k$. Equations of boundary motion (1) and (2) admit a pair of exact solutions in the form $\psi=\pm \hat H \eta$. These solutions describe the dispersion-free propagation of nonlinear waves in the direction of the external electric field or oppositely to it (depending on the sign at). The system energy is expressed in the following form:
$$H=H_0+H_1=-\frac{1}{2}\int \left(\psi \hat H \psi_x+\eta \hat H \eta_x \right)dx-\frac{1}{2}\int y\left([\hat H \psi_x^2-\psi_x^2]+[\hat H y_x^2-y_x^2]\right)dx,$$
where $H_0$ is the term corresponding to the linear approximation; $H_1$ corresponds to the quadratically nonlinear approximation.

To simulate system (1), (2), we take the initial conditions in the form
$$\eta(x,0)=\eta_1+\eta_2=a_1 \cos(k_1x)+a_2 \cos(k_2x),\quad \psi(x,0)=\hat H \eta_1-\hat H \eta_2.$$
In the linear approximation, these conditions correspond to two periodic waves with an amplitude $a_{1, 2}$ and wavelength $\lambda_{1,2}=k_{1, 2}$. The waves propagate in opposite directions.

Let us present results of the numerical solution of system (1), (2) for parameters $a_1 = 0.1,\, a_2 =0.05, \,k_1 = 1$, and $k_2 = 2$. The computations were carried out using pseudo-spectral methods of calculating spatial derivatives and integral operators with the total number of harmonics $N = 2^{15} = 32 768$. Time integration was carried out by the explicit fourth-order Runge-Kutta method with step $dt = 5 \cdot 10^{-7}$. To stabilize the numerical scheme, higher harmonics with
wavenumber $k\geq 10 \,923$ were equated to zero at every step of the time integration. Note that the accuracy of the numerical methods used is high. By the end of the calculation interval, the relative change in the total energy constituted a small quantity $\Delta H/H \sim 10^{-9}$.

The simulation results testify about the formation of weak singularities at the fluid boundary. Figure 1 shows the shape of the fluid boundary, as well as steepness
$\eta_x$ and curvature $K=\eta_{xx}/(1+\eta_x^2)^{3/2}$ of the surface. The observed behavior of the system is very close to the process of the formation of root singularities at the free boundary of the fluid with the development of hydrodynamic instabilities [1-4]. One can see from Fig. 1 that a region ($x_0\approx2.73$) with a sharp drop of curvature has formed at the boundary by the end of the calculation interval. The slope angles remain small and the maximum amplitude of the boundary curvature increases approximately by 30 times as compared to the initial value.

\begin{figure}[h]
\center{\includegraphics[width=0.7\linewidth]{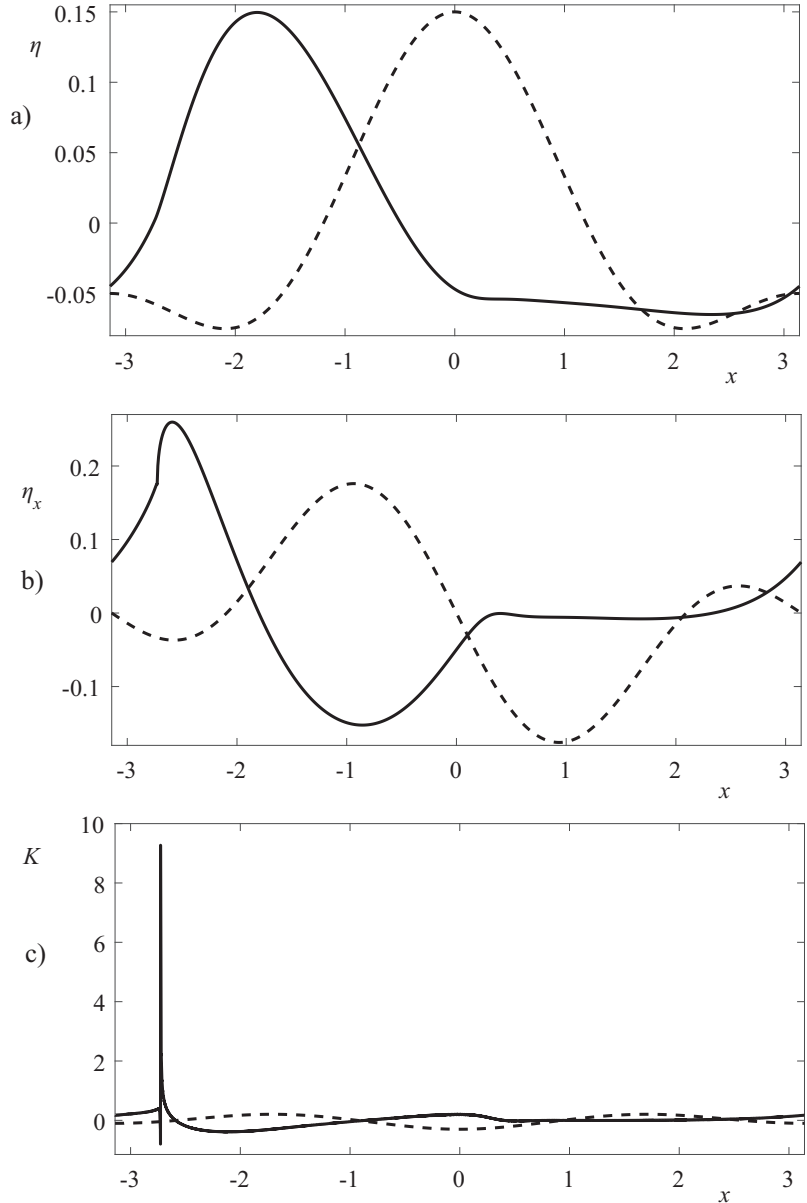}}
\caption{(a) Shape of the fluid boundary and (b) steepness and (c) curvature of the surface at the initial time instant (dashed lines) and in the end of calculation interval $t \approx 125.3$ (solid lines).}
\label{buble}
\end{figure}

The singular behavior of the system is also suggested by the spectral functions $|\eta_k|^2$ presented at successive time instants in Fig. 2. For a smooth and continuous function, the spectrum must decrease by an exponential law. Figure 2 shows that the surface spectrum attenuates by the end of the calculation interval
($t_0\approx 125.3$) not exponentially but by a power law:
$$ |\eta_k|^2\sim k^{-5},$$
which unambiguously points to the formation of a singularity. As a consequence, the boundary curvature in the Fourier space also acquires a power dependence: $|K_k|\sim k^{-1/2}$.

\begin{figure}[h]
\center{\includegraphics[width=0.7\linewidth]{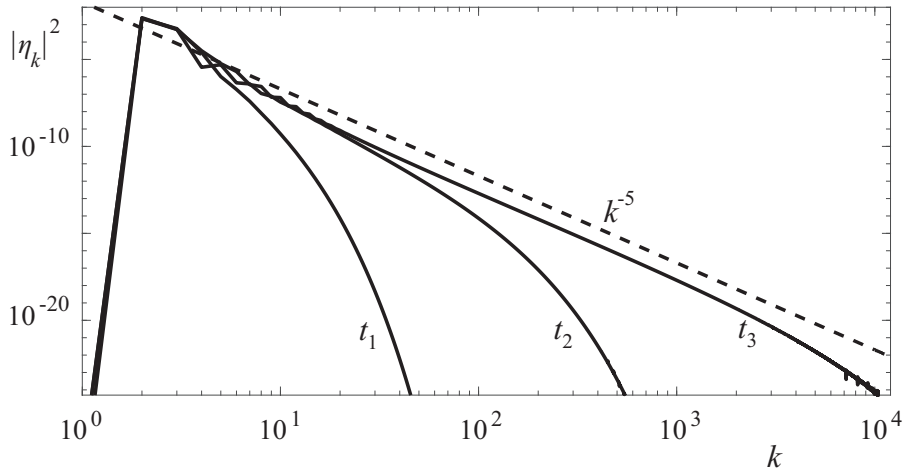}}
\caption{Spectral functions $|\eta_k|^2$ at successive time instants $t_1 = 15,\, t_2 = 110$, and $t_3 = t_0 = 125.3$. The dashed line corresponds
to power dependence $k^{-5}$.}
\label{buble}
\end{figure}

A distinctive feature of root singularities is universal self-similar behavior of the fluid boundary near the singularity. In particular, regardless of the physical nature [1-5], the curvature of the boundary near the singularity must be described by the power function
\begin{equation}\label{curv}K\approx \eta_{xx}\sim(x+x_0)^{-1/2},
\end{equation}
where $x_0$ is the coordinate of the singular point. Figure 3 shows the boundary curvature on the segment $-2.73 \leq x\leq -2.3$. The dashed line in the figure
corresponds to the root dependence (3). One can see that the presented functions almost coincide with each other on approaching point $x_0$. In contrast to dependence (3), the calculated curvature of the boundary is finite at point $x_0$. This fact may be related to the finite accuracy of the used numerical methods.

\begin{figure}[h]
\center{\includegraphics[width=0.8\linewidth]{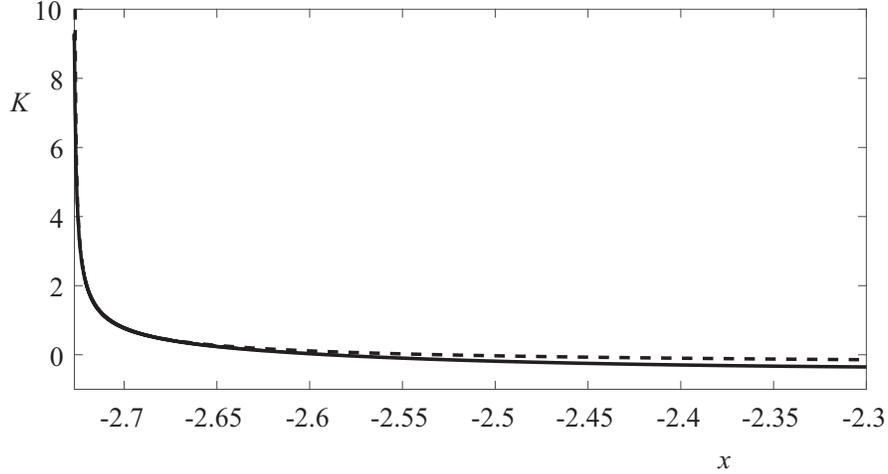}}
\caption{Curvature of the boundary at the time instant $t_0 =125.3$ (solid line) and power dependence $f(x) = 0.2(x -2.7268)^{-1/2} - 0.45$ (dashed line).}
\label{buble}
\end{figure}

Thus, numerical solution of system of equations (1), (2) demonstrates the formation of weak singularities at the fluid boundary in a finite time. The obtained data testify that root singularities appearing in different models of hydrodynamic instabilities [1-4] are the most probable candidates for this part.

It is important that the formation of singularities does not break the small-angle approximation and, potentially, can be described analytically. Apparently, the weak singularities observed in this work can go over into strongly nonlinear regions of shock fronts in which the field strength and fluid velocity undergo a discontinuity [20].

\section*{Acknowledgments}
I am deeply grateful to N.M. Zubarev for fruitful discussion of the results. This work was supported by the Russian Foundation for Basic Research (project no. 16-38-60002 mol\_a\_dk) and, in part, by state order topic 0389-2015-0023 and the Russian Foundation for Basic Research (projects nos. 17-08-00430 and 19-08-
00098).

\textbf{REFERENCES}

1. Moore D.W. // Proc. R. Soc. Lond. Ser. A. 1979. V. 365. P. 105.

2. Zubarev N.M., Kuznetsov E.A. // J. Exp. Theor. Phys. 2014. V. 119. P. 169.

3. Zubarev N.M. // J. Exp. Theor. Phys. 1998. V. 87. P. 1110.

4. Zubarev N.M. // Phys. Lett. A. 1998. V. 243. P. 128

5. Kuznetsov E.A., Spector M.D., Zakharov V.E. // Phys. Rev. E. 1994. V. 49. P. 1283.

6. Dyachenko S.A., Lushnikov P.M., Korotkevich A.O. // JETP Lett. 98, 675 (2013)

7. Taylor G. // Proc. R. Soc. Lond. A. 1964. V. 280. P. 383.

8. Kochurin E.A, Zubarev N.M. // Phys. Fluids. 2012. V. 24. P. 072101.

9. Zubarev N.M. // Phys. Fluids. 2006. V. 18. P. 028103

10. Karabut E.A., Zhuravleva E.N. // J. Fluid Mech. 2014. V. 754. P. 308.

11. Zubarev N.M., Karabut E.A. // JETP Lett. 2018. V 107. P. 412.

12. Melcher J.R., Schwarz W.J. // Phis. Fluids. 1968. V. 11. P. 2604.

13. Korovin V.M. // Tech. Phys. 2017. V. 62. P. 1316.

14.  Tao B., Guo D.L. // Comput. Math. Appl. 2014. V. 67. P. 627.

15. Tao B. // Comput. Math. Appl. 2018. V. 76. P. 799.

16. Zubarev N.M., Kochurin E.A. // J. Appl. Mech. Tech. Phys. 2013. V. 54. P. 212.

17. Zubarev N.M. // Phys. Lett. A. 2004. V. 333. P. 284.

18. Zubarev N.M.,Zubareva O.V. // Tech. Phys. Lett. 2006. V. 32. P. 886.

19. Zubarev N.M. // JETP Lett. 2009. V 89. P. 217.

20. Kochurin E.A. // J. Appl. Mech. Tech. Phys. 2018. V. 59. P. 79.

\begin{flushright}{ Translated by A. Nikol'skii }\end{flushright}

\end{document}